\documentstyle[pre,aps]{revtex} 

\begin{document}

\draft


\title
{ Diffusion-limited Reactions of hard-core Particles in one-dimension }

\author
{ P.-A. Bares and M. Mobilia}

\address
{
Institut de Physique Th\'{e}orique,
\'{E}cole Polytechnique F\'{e}d\'{e}rale de Lausanne,
CH-1015 Lausanne 
}

\date{ Revised October, 1998 }

\maketitle

\begin{abstract}

We investigate three different methods to tackle the problem 
of diffusion-limited reactions (annihilation) 
of hard-core classical particles in one
dimension. 
We first extend
an approach devised by 
Lushnikov and calculate for a single species
the asymptotic long-time and/or large distance
behavior
of the two-point correlation function. 
Based on a work by Grynberg et \it al.\rm , 
which was developed to treat stochastic
adsorption-desorption models, we provide in a second step the exact   
two-point correlation function (both for one and two-time) of Lushnikov's model. 
We then propose a new formulation of the
problem in terms of path integrals for pseudo-fermions. This formalism
can be used to advantage in the multi-species case, specially when applying perturbative
renormalization
group techniques.

\end{abstract}
\section{Introduction}

The recent interest in modelling low dimensional diffusion-limited reactions
has been stimulated in part by the experimental observation of anomalous 
kinetics
in low-dimensional systems \cite{Privman97}. The traditional approach to 
chemical 
reactions
is based on mean-field theory, i.e., rate equations for the densities of the
various reactants. The latter describe well the reaction kinetics in three
dimensions because diffusive transport of reactants allows to eliminate the
spatial
fluctuations of the concentrations. However, in lower dimensions, due to the
lack of phase space, 
the reactants spatial fluctuations can grow and develop 
inhomogeneities in the concentrations. 
Futhermore, even in the spatially homogeneous case, the rate equations are not
applicable in less than three dimensions; for example, in 
the two-species diffusion-limited annihilation the 
concentration of
the particles decays (for identical initial concentrations) slowlier than the
mean-field theory predicts. Thus
the fluctuation-dominated dynamics is beyond the classical theories, yet can be
accounted for by simple one-dimensional models of hard-core particles. 
The latter are solved
either numerically or analytically by applying techniques from  
(classical or quantum) statistical mechanics \cite{Privman97}. 
In particular, exact solutions
have been obtained by inter-particle distribution methods, by relating the
systems
to dual and solvable 1D models \cite{Glauber,Ra,Family,Droz} (Kinetic Ising and Potts) and 
by mapping the
diffusion-reaction processes to an imaginary-time dynamics 
of quantum spin chains with non-hermitian Hamiltonians \cite{Droz}. An alternative
and fruitful approach has been developed by Cardy and collaborators 
\cite{Cardy1,Cardy2,Cardy3,Lee}: 
the idea is to
reformulate the original problem in terms of a field theory of interacting 
bosons 
and subsequently use renormalization group techniques. This is 
a powerful method as it
applies to arbitrary dimension and low densities of particles, a regime where 
universal behavior (scaling) is usually observed.    
Despite the progress achieved in this field, the multi-species case
is still poorly understood. Furthermore, when the density of particles is high, 
the hard-core constraint on the dynamics of the diffusing particles becomes 
important.
Experimentally the single-species fusion model used to describe
the photoluminescence of excitons diffusing along one dimensional chains
is seen to apply only as long as the initial exciton densities are small \cite{Privman97}.
To our knowledge, there has been no systematic effort yet to investigate 
theoretically the regime
of high densities of reactants.\\
This is the first in a series of technical papers where 
we will explore a new 
approach
in an attempt to remedy the difficulties encountered so far. 
We propose to start from a quantum spin chain formulation of the master 
equation,
fermionize and subsequently
apply renormalization group techniques to deal with
the interaction terms arising in the multi-species case. Note that
in the single species case, the method
applies at arbitrary densities as the hard-core of the classical particles is
automatically accounted for by the Fermi statistics.
When various species react and diffuse, it is appropriate to distinguish two
cases: i) the various species have infinite on-site
repulsion with themselves only; this can be treated readily following the 
methods outlined in the following sections; ii) the particles of different 
species all have a hard core constraint; this is far more difficult and will be
investigated elsewhere. 
As with the other methods devised so far, 
there is a price to pay: the calculations involved are sometimes extremely
tedious.\\ 
The purpose of this first paper is modest as we will focus on the
single-species case for which fermion-fermion interactions do not arise as a consequence of mapping
classical particles to fermions.
We want to show explicitly how by elementary means we can
reproduce known results 
(the long-time behaviour of the density), and
derive some new results, i.e., the explicit and exact analytical form of the 
two-point correlation function of one and two-time.  
The paper is intended as an introduction to the fermionic functional
integration approach which we will combine in forthcoming papers
with the renormalization techniques to treat the multi-species case.\\
In the first section, we define the model, introduce notations and
review an extension of a method developed by Lushnikov. The second section is
devoted to the application of an elegant technique introduced 
by Grynberg et {\it al.} in a different context to evaluate the two-point (one-time) correlation function.
The third section
deals with a new and powerful formulation of the problem
in terms of path integrals of pseudo-fermionic
variables. We illustrate the technique by computing the two-time correlation function. 
A brief discussion concludes the paper.

\section{ Lushnikov's approach}
This section introduces notations, summarizes and extends
Lushnikov's genuine approach.
We consider a lattice of $N$ (even) sites (length $L=Na,\, a=1$ and 
assume $\frac{N}{2}$ even), with periodic boundary conditions, on which 
classical 
(spinless) particles with  hard-core can diffuse (annihilate)
to
adjacent empty (occupied) sites with rate $D$. Whenever the arrival site is
occupied, an annihilation-reaction ($1+1\rightarrow \O$)
takes place. A source of intensity $J$ injects pairs of particles on adjacent
sites ($\O\rightarrow 1+1$).
Lushnikov \cite{Lushnikov} has managed to rewrite the master equation that 
describes the
annihilation and diffusion processes described above in terms of 
an imaginary-time
Schr\"odinger equation:
\begin{eqnarray}
\label{1.4}
\displaystyle
\frac{d}{dt}|\psi (t)\rangle = {\mathcal L}|\psi (t)\rangle\;\; ,
\end{eqnarray}
where ${\mathcal L}$ denotes the so called ``Liouvillian",
which by abuse of language we will call
``non-hermitian Hamiltonian''. $|\psi (t)\rangle$
represents the state of the system at time t,
\begin{eqnarray}
\label{1.5}
\displaystyle
\ |\psi (t)\rangle =\sum_{\{n\}} \Bigr( P(\{n\},t)\prod_{m'(\{n\})} 
\sigma^+_{m'(\{n\})}
\Bigr)|0\rangle\;\;,
\end{eqnarray}
where $m'(\{n\})$ represents the sites of configuration $\{n\}$
which are occupied. The Liouvillian is given by \cite{Lushnikov} 
\begin{eqnarray}
\displaystyle
\label{1.7}
{\mathcal{L}}=(J+D)\sum_{m}\Big(\sigma_m^+\sigma_{m+1}^+ 
+\sigma_{m}^+\sigma_{m+1}^- + \sigma_{m}^-\sigma_{m+1}^+ +
\sigma_{m}^-\sigma_{m-1}^-\Big) +D\sum_{m} \Big(\sigma_{m+1}^-\sigma_{m}^- -
\sigma_{m+1}^+\sigma_{m}^+   -2\sigma_{m}^+\sigma_{m}^- \Big)-JN
\end{eqnarray}
When $J=0$, i.e., no source, in addition to diffusive processes with rate $D$, 
only irreversible reactions ($1+1\rightarrow \O$), with
rate $2D$ take place.\\
For finite ($J>0$) source, the diffusive processes ($1+\O\rightarrow \O +1$ and 
$\O + 1\rightarrow 1 +\O $) take place with rate $J+D$ and we also have  reversible reactions:
 particles are
annihilated ($1+1\rightarrow \O$) with rate $J+2D$ and created ($\O \rightarrow 1+1$)  with rate
$J$. It is worth emphasizing that these rates are not independent, and are chosen such that 
the Liouvillian is quadratic in 
the spin variables for a single
species (the higher order terms cancel due to the properties of Pauli matrices).
We also point out that in this model the ``annihilation rate" ($J+2D$) is always bigger 
to the ``creation rate" ($J$).
In the n-species case, this property does no longer hold if we assume
hard core repulsion between all species. Indeed, one obtains a spin
Hamiltonian ($S=n/2$) that is a polynomial of higher order in the 
spin operators and in general cannot be solved exactly. 
If however, we assume infinite on-site repulsion only bewteen
particles of the same species, we can rewrite the Hamiltonian as a quadratic
form of coupled spins $1/2$. The latter can be solved by the techniques
presented below.\\
To solve the Schr\"odinger equation in imaginary time, Lushnikov
performs a 
Jordan-Wigner transformation and introduces the fermionic operators
 $a_m =\prod_{j<m}(1-2 n_j)\sigma_{m}^- $.\\
Because of the form of the resulting non-hermitian Hamiltonian, it is
appropriate to work with Fourier modes $a_q=\frac{e^{i\frac{\pi}{4}}}{\sqrt{N}}\sum_{m} a_m e^{-iqm}$.
The antiperiodic boundary conditions \cite{Mattis} lead to:
$q=\pm (2l-1)\pi/N\;, \;l=1,2,\ldots,N/2$.
On Fourier transformation, the evolution operator reads
${\mathcal L}=\sum_{q>0} {\mathcal L}_q$, where ($n_q\equiv a_q^+a_q$)
\begin{eqnarray}
\label{1.14}
\displaystyle
{\mathcal L}_q=2(J+D) \Bigr \{\cos{q(n_{q}+n_{-q})}+\sin{q(a_{q}a_{-q}- 
a_{q}^+a_{-q}^+)}\Bigr \} 
 +2D \Bigr \{\sin{q(a_{q}a_{-q}+ a_{q}^+a_{-q}^+)}  -(n_{q}+n_{-q}) \Bigr 
\}-JN\;\;,
\end{eqnarray}
Now, by a BCS-like Ansatz, 
\begin{eqnarray}
\label{1.15}
\displaystyle
\ |\psi(t)\rangle=\prod_{q>0}|\psi_q (t)\rangle  =\prod_{q>0}\Bigr( A_q (t) 
a_q^+a_{-q}^+ + B_q
(t)\Bigr)|0\rangle,
\end{eqnarray}
Lushnikov \cite{Lushnikov} is able to decouple the dynamical equation as
\begin{eqnarray}
\label{1.16}
\displaystyle
\  \frac{d}{dt}|\psi_q (t)\rangle = {\mathcal L}_q|\psi_q (t)\rangle
\;,\;\forall q>0
\end{eqnarray}
For a lattice which is initially completely occupied, i.e., $A_q(0)=1, 
B_q(0)=0$, 
one solves the above equations by
\begin{eqnarray}
\displaystyle
\label{1.17}
\ A_{q}(t)=\frac{1}{4(J+2D)\sin^2{(\frac{q}{2})}}(p_{2}{\rm e}^{p_{2}t} - 
p_{1}{\rm e}^{p_{1}t}) \;\;,\;\; \ B_{q}(t)=\frac{-(J+2D)\sin{q}}{2(J+2D)\sin^2{(\frac{q}{2})}}({\rm e}^{p_{2}t} 
- {\rm e}^{p_{1}t})
\end{eqnarray}
where
\begin{eqnarray}
\displaystyle
\label{1.19}
\ p_{1}=-2(J+2D)(1-\cos{q}) \;\;,\;\; \ p_{2}=2J (1+\cos{q})
\end{eqnarray}
In the absence of source ($J=0$), the solution simplifies considerably to
\begin{eqnarray}
\displaystyle
\label{1.20}
\ A_{q}(t)=\exp(-4Dt(1-\cos{q})) \;\;,\;\; B_{q}(t)=\cot{(\frac{q}{2})}\Bigr(\exp(-4Dt(1-\cos{q})-1)\Bigr)
\end{eqnarray}
At this point it is worth noting that the ket
$|\psi(t)\rangle$ 
characterizes the state of the system at any time without however being an 
eigenvector 
of ${\mathcal L}$ (The Liouvillian is not normal, however see below).\\ 
In ref.\cite{Lushnikov}, Lushnikov calculates the density of particles 
by the method of the
generating function, which we extend in order to evaluate the two-point
correlation function.
The density reads
\begin{eqnarray}
\displaystyle
\label{2.1}
\  \rho(t) = \sum_{\{n\}} \widetilde n_i(\{n\})\,\, P(\{n\},t), \forall i,
\end{eqnarray}
Similarly, the two-point correlation function is written as 
\begin{eqnarray}
\displaystyle
\label{2.2}
\ {\mathcal G}_r(t)\equiv\langle\widetilde n_i \widetilde n_{i+r} \rangle (t)= 
\sum_{\{n\}} \widetilde n_i(\{n\})\widetilde n_{i+r}(\{n\})\,\, P(\{n\},t) \,,
\end{eqnarray} 
where the translational symmetry of the system has been used.
We observe that in  
this formalism  
\begin{eqnarray}
\displaystyle
\label{2.4}
\ \rho (t)= \langle 0|\exp(\sum_n \sigma_n^-)\,\, n_i |{\psi}(t)\rangle \;\;,\;\;
{\mathcal G}_r(t) =\langle 0| \exp(\sum_n \sigma_n^-)\,\, n_i n_{i+r}  
|{\psi}(t)\rangle
\end{eqnarray}
as one can check using the explicit form of $|\psi(t)\rangle$ (develop the 
exponential, order
each term and perform a Jordan-Wigner transformation 
\cite{Lushnikov}).
It is appropriate to consider the following generating function:
\begin{eqnarray}
\displaystyle
\label{2.6}
\ G(x,y,z,t)= \langle 0| exp \Bigr(x \sigma_i^- + y\sigma_{i+r}^- + 
z \sum_{n\neq i,i+r}\sigma_n^-  \Bigr)|\psi(t)\rangle= \sum_{\{n\}}x^{\widetilde 
n_i} y^{\widetilde
n_{i+r}}z^{\sum_{j}\widetilde n_j - \widetilde n_i - \widetilde n_{i+r}} P(\{n\},t)
\end{eqnarray}
So we have,
\begin{eqnarray}
\displaystyle
\label{2.7}
 \rho(t)=\frac{\partial}{\partial x}G(x,z,z,t)\Big|_{x,y,z=1}\;\;,\;\;
 {\mathcal G}_r(t)=\frac{\partial^2}{\partial x \partial y}G(x,y,z,t)
 \Big|_{x,y,z=1}
\end{eqnarray}
To compute the generating function, we rewrite the state as
\begin{eqnarray}
\displaystyle
\label{2.9}
|\psi(t)\rangle=\prod_{q>0}\Bigr( A_q (t) a_q^+a_{-q}^+ + B_q(t)\Bigr)|0\rangle=
\prod_{q>0}\Bigr( B_q(t)-\frac{2A_q(t)}{N}\sum_{n>m} a_m^+a_n^+ 
\sin{q(n-m)}\Bigr)|0\rangle
\end{eqnarray}
Note the normalisation condition due to the conservation of probability:
\begin{eqnarray}
\displaystyle
\label{2.10}
\sum_{\{n\}}P(\{n\},t)=\prod_{q>0}\Big(B_q(t)-A_q(t) \cot{\frac{q}{2}}\Big)=1
\end{eqnarray}
Next expand the argument of the exponential as 
\begin{eqnarray}
\displaystyle
\label{2.11}
G(x,y,z,t)=\langle 0| \Bigg(\openone+(x a_i+y a_{i+r} + z\sum_{n\neq i,i+r}a_n) 
+
\Big\{z^2\sum_{i+r\neq n>n'\neq i}a_n a_{n'} + 
xz(a_i\sum_{n<i}a_n+\sum_{n>i,n\neq i+r}a_n a_i)+\nonumber\\
+yz(\sum_{n>i+r}a_n a_{i+r}+ a_{i+r}\sum_{
n<i+r, n\neq i} a_n ) +xy a_{i+r}a_i
     \Big\}+\ldots \Bigg) 
     \prod_{q>0} \Biggr( 2\frac{A_q(t)}{N} \sum_{n>m} a_m^+a_n^+ \sin{q(n-m)}  -
     B_q(t)\Biggr)  |0\rangle
\end{eqnarray}

In this expression, only the terms proportional to ``$xy$" contribute to  
${\mathcal G}_r(t)$. Let us call the first of these terms  $G_1$,
\begin{eqnarray}
\displaystyle
\label{2.12}
G_1=\langle 0| xy a_{i+r}a_i
     \sum_{q>0} \Biggr\{ 2\frac{A_q(t)}{N} \sum_{n>m} a_m^+a_n^+
     \frac{\sin{q(n-m)}}{A_q(t)\cot{\frac{q}{2}}-B_q(t)} \prod_{q\neq
     q'>0} \Biggr(\frac{B_{q'}(t)}{B_{q'}(t)-A_{q'}(t)\cot{\frac{q'}{2}}}\Biggr) 
       \Biggr\}  |0\rangle = \nonumber\\ =xy 
\frac{2}{N}\Biggr(\sum_{q>0}\frac{\sin{qr}}{\cot{\frac{q}{2}}-
       \frac{B_q(t)}{A_q(t)}}\Biggr)\Biggr(\prod_{q\neq 
q'>0}\frac{1}{1-\frac{A_{q'}(t)}{B_{q'}(t)}\cot{\frac{q}{2}}  } \Biggr)
\end{eqnarray}
In the absence of source, we have $A_q(t)\rightarrow 0$ and
$B_q(t)\rightarrow -\cot{\frac{q}{2}} $ exponentially fast (see \ref{1.20}), so 
that in the thermodynamic limit $(N\rightarrow
\infty)$, the asymptotic behaviour of ${\mathcal G}_r(t)$ follows as
\begin{eqnarray}
\displaystyle
\label{2.13}
{\mathcal G}_r(t)\sim \frac{\partial ^2}{\partial x \partial y}G_1 \rightarrow 
\frac{1}{\pi} \int_{0}^{\pi} \frac{dq 
\sin{qr}}{\cot{\frac{q}{2}}-\frac{B_q(t)}{A_q(t)}}
\end{eqnarray}
We can do the same for the density and in the thermodynamic limit, 
one obtains
\begin{eqnarray}
\displaystyle
\label{2.15}
\rho(t)=\frac{\partial}{\partial x}  \prod_{q>0} \langle 0| 
\Biggr(\openone+\Big(x a_i + z\sum_{n\neq i}a_n \Big) +
\Big\{z^2\sum_{i\neq n,n', n> n'}a_n a_{n'} + 
xz(a_i\sum_{n<i}a_n+\sum_{n>i} a_n a_i) + \ldots \Big \}  \Biggr) \times 
\nonumber\\ 
     \times \Biggr( 2\frac{A_q(t)}{N} \sum_{n>m} a_m^+a_n^+ \sin{q(n-m)}  -
     B_q(t)\Biggr)  |0\rangle\Big|_{z=1,x=0}
     \rightarrow \frac{1}{\pi}\int_{0}^{\pi}\frac{dq}{1-\frac{B_q(t)}{A_q(t) 
\cot{\frac{q}{2}}}}
\end{eqnarray}
In the above, we used the following identities:
\begin{eqnarray}
\label{2.16}
\displaystyle
 &\sum_{m>i}&\sin{q(m-i)}=\frac{1}{2}\Bigr(\cot{\frac{q}{2}}+
\frac{\cos{q(i-\frac{1}{2})}}{\sin{\frac{q}{2}}}\Bigr)\;\;\;,\;\;\;
\sum_{m<i}\sin{q(m-i)}=\frac{1}{2}\Bigr(\cot{\frac{q}{2}}-\frac{\cos{q(i-\frac{1
}{2})}}{\sin{\frac{q}{2}}}\Bigr)\nonumber \\ &\Rightarrow & \sum_{n>m}\sin{q(n-m)}=\frac{N}{2} \cot{\frac{q}{2}}
\end{eqnarray}
The evaluation of the two-point correlation function  
requires the calculation of all the terms proportional to
$xy$, which in general is a very hard task. 
In the following sections we will be able 
to solve this
difficulty by reformulating the problem in a different language.\\
Using the explicit form of $A_q(t)$ and $B_q(t)$ (\ref{1.17}-\ref{1.20}) and the 
results of appendix A, we find the
asymptotic behaviour of the density in the
({\it irreversible}) {\it critical case} as \cite{Lushnikov}
\begin{eqnarray}
\label{2.19}
\displaystyle
\rho(t)=\frac{e^{-4Dt}}{\pi}\int_{0}^{\pi} dq \, e^{4Dt
\cos{q}}=e^{-4Dt}I_0(4Dt)\sim\frac{1}{\sqrt{8\pi Dt}}
\end{eqnarray}
Similarly the asymptotic
behaviour of the two-point correlation function is 
\begin{eqnarray}
\label{2.20}
\displaystyle
\ {\mathcal G}_r (t)&\sim & \frac{e^{-4Dt}}{\pi} \int_{0}^{\pi} dq 
\frac{\sin{qr}}{\sin{q}}
(1-\cos{q})e^{4Dt\cos{q}} \nonumber\\ 
&=& e^{-4Dt}\sum_{0\leq n <r}\Bigr\{ I_{2n-r+1}(4Dt) - I_{2n-r} (4Dt) \Bigr\}  \sim \frac{\pi r}{(8 \pi Dt)^{\frac{3}{2}}}
\end{eqnarray}
which implies
\begin{eqnarray}
\label{2.21}
\displaystyle
C_r(t)\sim  -\frac{1}{8\pi Dt}
\end{eqnarray}
Unfortunately, in the massive case (when the source intensity is finite), 
this method applies only to the computation of the
density. Assuming 
$Dt,Jt \gg 1$, we find
\begin{eqnarray}
\label{2.22}
\displaystyle
\rho(t)=\rho_{eq}+2(J+2D)\int_{t}^{\infty} dt'
 e^{-4(J+D)t'}\Big(I_0(4Dt') -I_1(4Dt')\Big)  \\ 
\sim\frac{\sqrt{J}}{\sqrt{J}+\sqrt{J+2D}}+\Big(1+\frac{J}{2D}\Big)
\frac{e^{-4Jt}
}{8Jt\sqrt{8\pi Dt}}
\end{eqnarray}
where $\rho_{eq}= \frac{\sqrt{J}}{\sqrt{J}+\sqrt{J+2D}}$ represents 
the equilibrium
value of the density, in agreement with Lushnikov's result \cite{Lushnikov}.
In the next section we provide the full two-point correlation function
when $J>0$.
\section{The pseudo-fermionic approach}
In this section we evaluate the full two-point correlation function 
in the general case by means of a powerful
formalism. The central idea is to perform on the fermionic non-hermitian (and
non-normal)
Hamiltonian
a  generalized Bogoliubov transformation which allows us to work with a diagonal evolution operator
(see Ref.\cite{Stinch1,Stinch2}).
Following  previous works \cite{Stinch1,Stinch2,Schutz,Droz}, we denote each of the $2^N$ possible configurations  
  by a ket $|n\rangle $:
\begin{eqnarray}
\label{3.1}
\displaystyle
\langle n|n'\rangle = \delta_{n,n'}\;\;\;,\;\;\;
\sum_{n}|n\rangle \langle n|= \openone
\end{eqnarray}
In this Fock space, we can efficiently record
the probabilities for the various configurations in the ket

\begin{eqnarray}
\label{3.2}
\displaystyle
|P(t)\rangle\equiv \sum_{n} P(n,t)|n\rangle
\end{eqnarray}
The master equation governing the
dynamics of the annihilation and diffusion processes 
described in the previous section
can be rewritten as
\begin{eqnarray}
\label{3.3}
\displaystyle
\frac{\partial}{\partial t}|P(t)\rangle={\cal U}|P(t)\rangle =
\sum_{n}\partial_t P(n,t)|n\rangle =\sum_{n,n'}\Big(A(n'\rightarrow n) P(n',t)
- A(n\rightarrow n') P(n,t)\Big)|n\rangle
\end{eqnarray}
where ${\cal U}$ denotes the evolution operator, $A(n'\rightarrow n)$ 
and $A(n\rightarrow n')$
represent the transition rates $J$ and $D$ in Lushnikov's formulation. 
The matrix
elements for the operator ${\cal U}$ are 
\begin{eqnarray}
\label{3.4}
\displaystyle
\langle n'|{\cal U}|n\rangle = A(n \rightarrow n'), \forall n'\neq n \\
\langle n|{\cal U}|n\rangle =-\sum_{n'\neq n} A(n \rightarrow n')
\end{eqnarray}
This Fock-space formulation was used in Ref. \cite{Stinch1,Stinch2}
to study a stochastic  adsorption-desorption problem.
In what follows, we will specifically focus on the  reaction-diffusion 
problem in one dimension.\\
Let us now introduce the left and right steady-states, respectively 
\begin{eqnarray}
\label{3.5}
\displaystyle
\langle\widetilde\chi| \equiv \sum_n \langle n|\;\;\;,\;\;\;|\chi\rangle \equiv \sum_n P(n,eq)|n\rangle
\end{eqnarray}
where $P(n,eq)$ denotes the probability for a configuration $|n\rangle$ at
equilibrium. It is easy to check that
$e^{{\cal L}t}$ has no effect on $ |\chi \rangle$ and $\langle \widetilde \chi|$, 
and the
conservation of probability leads to $\langle \widetilde \chi|\chi \rangle 
=1$.\\
The transition probability from a configuration $|n\rangle$ to $|n'\rangle$ 
is simply: $W_{n,n'}(t)\equiv \langle n'| e^{{\cal L}t}|n\rangle$.\\
We intend to calculate the density and two-point one-time correlation functions 
of  a system initially in
the state $|\phi_0\rangle \equiv \sum_{n}P(n,t=0)|n\rangle$.
The occupation number operator $n_r$ being diagonal in the basis 
$\{|n\rangle\}$, we have 
\begin{eqnarray}
\label{3.7}
\displaystyle
\rho(t)=\sum_{n,n'}\langle n'|n_r|n'\rangle W_{n,n'}(t) P(n,t=0)=
\sum_{n,n'}\langle n'|n_r e^{{\cal L}t}|n\rangle  P(n,t=0)=
\langle \widetilde \chi| n_r e^{{\cal L}t}|\phi_o\rangle
\end{eqnarray}
and similarly,
\begin{eqnarray}
\label{3.8}
\displaystyle
{\cal G}_r(t)=\sum_{n,n'}\langle n'|n_{l} n_{m}|n'\rangle W_{n,n'}(t) P(n,t=0)=
\sum_{n,n'}\langle n'|n_{l} n_{m} e^{{\cal L}t}|n\rangle  P(n,t=0)=
\langle \widetilde \chi|n_{l} n_{m} e^{{\cal L}t}|\phi_o\rangle, \, 
\end{eqnarray}
where $r=|m-l|$.
At this point, we perform a  generalized Bogoliubov 
transformation (rotation supplemented by a rescaling),
\begin{eqnarray}
\label{3.9}
\displaystyle
 \left(
 \begin{array}{c}
 \xi_q^+  \\
 \xi_{-q} 
\end{array}\right)\
=
\left(
 \begin{array}{c c}
 \alpha \cos{\theta_q} & \alpha^{-1} \sin{\theta_q}  \\
  -\alpha \sin{\theta_q} & \alpha^{-1} \cos{\theta_q}
\end{array}\right)\
\left(
 \begin{array}{c}
 a_q^+  \\
 a_{-q} 
\end{array}\right)\
\end{eqnarray}
in order to obtain a diagonal representation for the evolution operator.\\
This transformation is orthogonal, i.e., invertible 
and  canonical, in the sense 
that it preserves the anticommutation relations of the
$a_q$'s, namely:
\begin{eqnarray}
\label{3.10}
\displaystyle
\{\xi_q^+,\xi_{q'}\}=\delta_{q,q'}\,, \{\xi_q^+,\xi_{q'}^+\}=\{\xi_q,\xi_{q'}\} 
=0
\end{eqnarray}
Despite the fact that the $\xi_q$ 
and $\xi_q ^+$ {\it are
not adjoint} of each other, this representation will be very useful in 
the following.\\
We set $\alpha=\Big(\frac{J}{J+2D}\Big)^{\frac{1}{4}}$, 
so that the mode $q$ evolution operator becomes
\begin{eqnarray}
\label{3.12}
\displaystyle
-{\cal L}_q=2\Big(D(1-\cos{q})-J\Big)\Big((\xi_q^+\xi_q + \xi_{-q}^+\xi_{-q})
+\sin{\theta_q}^2(\xi_{-q}\xi_{-q}^+ + \xi_q \xi_q^+) +2(\xi_{-q}^+\xi_{q}^+ + 
\xi_q \xi_{-q})\Big)
\nonumber\\+\sqrt{J(J+2D)}\sin{q}\Big(\cos{2\theta_q} (\xi_{-q}^+\xi_{q}^+ + 
\xi_q \xi_{-q})
+ \sin{2\theta_q}(\xi_{q}\xi_{q}^+ + \xi_{-q}^+ \xi_{-q})  \Big) +2J
\end{eqnarray}
To get rid of the terms that do not conserve the number of  
pseudo-particles, we choose $\theta_q$ as
\begin{eqnarray}
\label{3.13}
\displaystyle
\tan{2\theta_q}=\frac{\sqrt{J(J+2D)}\sin{q}}{(J+D)\cos{q}-D}
\end{eqnarray}
so that the Hamilton operator becomes
\begin{eqnarray}
\label{3.16}
\displaystyle
{\cal L}= -\sum_{q>0}\lambda_q \Big(\xi_q^+ \xi_q + \xi_{-q}^+ 
\xi_{-q}\Big)=-\sum_{q} \lambda_q  \xi_q^+ \xi_q
\end{eqnarray}
where
\begin{eqnarray}
\label{3.15}
\displaystyle
\lambda_q= 2\Big(D(1-\cos{q})+J\Big)
\end{eqnarray}
on account of the periodic boundary conditions ($\sum_{q>0} 
\cos{q}=0$).
Now it is readily seen that $\langle 
\widetilde \chi|$ and 
$|\chi\rangle$ act, respectively, as left and right vacua, i.e.,
$\xi_q|\chi\rangle=0$ and 
$\langle \widetilde \chi|\xi_q^+ =0$.
To simplify the calculations, we
express the initial ket-state $|\phi_0\rangle$ in terms of the 
steady state 
$|\chi\rangle$.
We consider here two kinds of initial conditions:\\
i) The  whole lattice is initially filled.
We write $|\phi_0\rangle=|all\rangle$ and immediately
conclude $a_q^+|all\rangle =0$.
Using the inverse of (\ref{3.9}), one can check that
\begin{eqnarray}
\label{3.18}
\displaystyle
|all\rangle = \prod_{q>0}\Big(1-\cot{\theta_q}\xi_q^+ \xi_{-q}^+ 
\Big)|\chi\rangle =
\exp{\Big(-\sum_q\frac{\cot{\theta_q}}{2} \xi_q^+ \xi_{-q}^+}\Big)|\chi\rangle
\end{eqnarray}
ii) The lattice is initially empty; one can check in the same way
\cite{Stinch2} that :
\begin{eqnarray}
\label{3.19}
\displaystyle
|\phi_0\rangle=|0\rangle = \prod_{q>0}\Big(1+\tan{\theta_q}\xi_q^+ 
\xi_{-q}^+\Big)|\chi\rangle =
\exp{ \Big(\sum_q\frac{\tan{\theta_q}}{2} \xi_q^+ \xi_{-q}^+}\Big)|\chi\rangle
\end{eqnarray}
The time dependence of $\xi_k(t)$ and $\xi_k^+(t)$ follow as
\begin{eqnarray}
\label{3.20}
\displaystyle
\xi_k(t)=e^{-{\cal L}t} \xi_q e^{{\cal L}t} = e^{-\lambda_q t} \xi_k \\
\xi_k^+(t)=e^{-{\cal L}t} \xi_q^+ e^{{\cal L}t} = e^{\lambda_q t} \xi_k^+ \\
\end{eqnarray}
Futhermore, we have 
\begin{eqnarray}
\label{3.21}
\displaystyle
\langle \xi_{k_1}\xi_{k_2}\rangle (t=0)\equiv \langle \widetilde 
\chi|\xi_{k_1}\xi_{k_2}|all\rangle=
\cot{\theta_{k_1}}\delta_{k_1,-k_2}
\end{eqnarray}
and
\begin{eqnarray}
\label{3.22}
\displaystyle
\langle \xi_{k_1}\xi_{k_2}\xi_{k_3}\xi_{k_4}\rangle (t=0)=
\cot{\theta_{k_1}}\cot{\theta_{k_3}}\delta_{k_1,-k_2}\delta_{k_3,-k_4}+
\cot{\theta_{k_1}}\cot{\theta_{k_2}}\Big(\delta_{k_1,-k_4}\delta_{k_2,-k_3} -   
\delta_{k_1,-k_3}\delta_{k_2,-k_4} \Big)
\end{eqnarray}
as one can check by applying Wick's theorem (see also  section {\rm IV}).
Using the properties of Fourier transform and the generalized Bogoliubov 
transformation 
(\ref{3.9}), the expression of the density and the two-point
correlation function become respectively (for a lattice initially  
filled, $|\phi_0\rangle=|all\rangle$):
\begin{eqnarray}
\label{3.23}
\displaystyle
\rho (t)= 
\frac{1}{N}\sum_{k} \sin^2{\theta_k} -
 \sum_{k,k'}\frac{e^{i(k'-k)l}}{N}\sin{\theta_k} \cos{\theta_{k'}} 
 \langle \widetilde \chi|\xi_{-k}\xi_{k'}e^{{\cal L}t}|all\rangle
\end{eqnarray}
and one can do the same for unconnected one-time correlation function 
${\cal G}_r (t)$ (\ref{3.8}).\\
To derive tractable formulas, we have performed tedious but straightforward 
calculations. Indeed, we have
extracted the time-dependence of $\rho(t)$ and  ${\cal G}_r (t)$
using (\ref{3.20}) and
commuted all the pseudo-creation operators to the
left of the pseudo-annihilation operators. In the expression 
$\rho(t)$ and ${\cal G}_r (t)$ 
only terms like $\langle\widetilde \chi|\xi_{-k}\xi_{k'}|all\rangle=\langle 
\xi_{-k}\xi_{k'}\rangle (t=0)$
and $\langle\widetilde \chi|\xi_{-k}\xi_{k'} \xi_{-q}\xi_{q'} |all\rangle=\langle 
\xi_{-k}\xi_{k'}\xi_{-q}\xi_{q'} \rangle (t=0)$ survive:
these were evaluated with the help of (\ref{3.21}, \ref{3.22}).\\
In the thermodynamic limit, we arrive at
\begin{eqnarray}
\label{3.25}
 \rho(t)&=&\frac{2}{N}\sum_{k>0} \Big(\sin^2{\theta_k}+ e^{-2\lambda_k t} 
\sin{\theta_k}\cos{\theta_k}\cot{\theta_k}\Big) \nonumber \\
 & \rightarrow & \frac{1}{\pi}\int_{0}^{\pi} dq \Big(\sin^2{\theta_q}+ 
e^{-2\lambda_q t}\cos^2{\theta_q}\Big)  
 = \frac{\sqrt{J}}{\sqrt{J}+\sqrt{J+2D}}+2(J+2D)\int_{t}^{\infty} dt' 
\Big(I_0(4Dt')-I_1(4Dt')\Big)e^{-4(J+D)t'}
 \end{eqnarray}
which coincides with (\ref{2.22}).\\
It is worth emphasizing that this result is general and works for both, the 
{\it massive} ($J\neq 0$) and the {\it critical} ($J=0$) cases.
The point here is that the limit $J\rightarrow 0$ is not singular, despite
the divergence of $\cot\theta\rightarrow-\infty$. In fact, integration over 
$k$ and $k'$ of terms
proportional to $\sin{\theta_k}\cot{\theta_{k'}}$ yields finite results. 
Therefore, we can perform the computations (\ref{3.9}) at $J$ finite and set
subsequently $J=0$ in  
$\rho(t)$ and ${\cal G}_r(t)$. \\
Similarly, the two-point correlation function (of one-time) is evaluated
as ($ r=|m-l|$):
\begin{eqnarray}
\label{3.26}
\displaystyle
{\cal G}_r(eq)=\rho_{eq}^2+\frac{1}{\pi^2}\Big(\int_{0}^{\pi} dq \,
\sin^2{\theta_q}\cos{qr}\Big)\Big(\int_{0}^{\pi}dq' \,
\cos^2{\theta_{q'}}\cos{q'r}\Big) +\Big(\frac{1}{2\pi}\int_{0}^{\pi} 
 dq \,\sin{2\theta_q}\sin{qr}\Big)^2 
 \end{eqnarray}
\begin{eqnarray}
\label{3.27.0}
\displaystyle
{\cal G}_r(t)-{\cal G}_r(eq)&=&\Big(\rho(t)^2 - \rho_{eq}^2\Big)+ 
\frac{1}{\pi^2}\Big(\int_{0}^{\pi}dq \, e^{-2\lambda_q
t}\cos^2{\theta_q}\cos{qr}\Big)\Big( \int_{0}^{\pi} dq' \, 
\cos^2{\theta_{q'}}\cos{q'r}\Big)
\nonumber\\ &+& \frac{1}{2\pi^2}\Big(\int_{0}^{\pi} dq \,
\sin{2\theta_q}\sin{qr}\Big)
\Big(\int_{0}^{\pi}dq'\, e^{-2\lambda_{q'}t}\cos^2{\theta_{q'}}\cot{\theta_{q'}}\sin{q'r}\Big)\nonumber\\
&-& \frac{1}{\pi^2}\Big(\int_{0}^{\pi} dq \, \sin^2{\theta_q}\cos{qr}\Big)
\Big(\int_{0}^{\pi}dq'\, e^{-2\lambda_{q'}t}\cos^2{\theta_{q'}}\cos{q'r}\Big)\nonumber\\
&-& \frac{1}{4\pi^2}\Big(\int_{0}^{\pi} dq \, \sin{2\theta_q}\sin{qr}\Big)
\Big(\int_{0}^{\pi}dq' \, e^{-2\lambda_{q'}t}\sin{2\theta_{q'}}\sin{q'r}\Big)\nonumber\\
&-& \Big( \frac{1}{\pi}\int_{0}^{\pi}dq \, e^{-2\lambda_qt}\cos^2{\theta_q}\cos{qr}\Big)^2
-\Big(\frac{1}{\pi^2} \int_{0}^{\pi} dq  \, e^{-2\lambda_{q}t}\cos^2{\theta_{q}}\cot{\theta_{q}}\sin{qr}  \Big)
 \nonumber\\ 
&\times & \Big( \int_{0}^{\pi}dq  \, e^{-2\lambda_{q}t}\sin^2{\theta_{q}}\cot{\theta_{q}}\sin{qr}  \Big)\;\; ,
 \end{eqnarray}
where we separated the static contribution to the correlation function from the
dynamic one. Using known properties of modified Bessel functions given (see appendix)  
and writing  $I'_m(x)\equiv \frac{d}{dx} I_m (x)$, we finally infer 
(notice that ${\cal G}_r(eq)=\rho_{eq}^2$) 
\begin{eqnarray}
\label{3.27.1}
\displaystyle
{\cal G}_r(t) &=& \rho(t)^2 + 2(J+2D) (A_0 - C_0) 
\int_{t}^{\infty}dt' e^{-4(J+D)t'}\Big\{I_r(4Dt')-I'_r(4Dt')\Big\} \nonumber\\ 
&+& \frac{\sqrt{J(J+2D)}B_0}{2}\int_{t}^{\infty}dt' e^{-4(J+D)t'} \frac{r}{2Dt'} I_{r}(4Dt')
 \nonumber\\ 
 &-&  \frac{\Big(J+2D\Big)^{\frac{3}{2}}}{\sqrt{J}}\sum_{0\leq n<r}\Big(\int_{t}^{\infty} dt'\, 
e^{-4(J+D)t'}\{2I_{2n-r+1}(4Dt')- 2I_{2n-r}(4Dt')
+I'_{2n-r}(4Dt')\}\Big) \nonumber\\
&-& \Big(2(J+2D)\int_{t}^{\infty} dt'\, e^{-4(J+D)t'} \Big\{I_r(4Dt')-I'_r(4Dt')\Big\}\Big)^2 \nonumber\\
&+& 2(J+2D)^2 \Big(\sum_{0\leq n<r}(2n-r)\Big(\int_{t}^{\infty} dt'\, 
e^{-4(J+D)t'}\frac{I_{2n-r}(4Dt')}{2Dt'} \Big)\Big) \nonumber\\
& \times & \Big(\sum_{0\leq n<r}\Big(\int_{t}^{\infty} dt'\, 
e^{-4(J+D)t'}\{2I_{2n-r+1}(4Dt')- 2I_{2n-r}(4Dt')
+I'_{2n-r}(4Dt')\}\Big)\Big)
 \end{eqnarray}
In the above formula $A_0$, $B_0$, $C_0$ have been defined by:
\begin{eqnarray}
\label{3.28}
\displaystyle
A_0\equiv \frac{1}{\pi}\int_{0}^{\pi} dq\,\cos^2{\theta_q}\cos{qr},\;\;
B_0\equiv  \frac{1}{\pi} \int_{0}^{\pi}dq\, \sin{2\theta_q}\sin{qr},\;\;
C_0 \equiv \frac{1}{\pi} \int_{0}^{\pi}dq\, \sin^2{\theta_q}\cos{qr} 
\end{eqnarray}
or more explicitly, in the massive case (with help of (\ref{A.16})),
\begin{eqnarray}
\label{3.28.bis}
\displaystyle
A_0 &=& \frac{(J+2D)}{\pi}\int_{0}^{\pi}\frac{dq}{\lambda_q} 
\,\Big(\cos{qr}-\frac{\cos{q(r+1)}+\cos{q(r-1)}}{2}\Big)= 
-(J+2D)\zeta^{r-1}\frac{(1-\zeta)^2}{4\sqrt{J^2+2JD}},\\
B_0 &=& \frac{\sqrt{J(J+2D)}}{\pi}  \int_{0}^{\pi}\frac{dq}{\lambda_q}
\,\Big(\cos{q(r+1)}-\cos{q(r-1)}\Big)  = 
-\zeta^{r-1}\frac{(1-\zeta^2)}
{2}, \\
C_0 &=& \frac{J}{\pi} \int_{0}^{\pi}\frac{dq}{\lambda_q}
\,\Big(\cos{qr}+\frac{\cos{q(r+1)}+\cos{q(r-1)}}{2}\Big)  
=J\zeta^{r-1}\frac{(1+\zeta)^2}{4\sqrt{J^2+2JD}}
\end{eqnarray}
with the notation 
\begin{eqnarray}
\label{3.29}
\displaystyle
\zeta\equiv \frac{D}{(J+D)+\sqrt{J^2+2JD}}\leq 1
\end{eqnarray}
Note that the above result, though obtained by elementary
means, is new. The formula can be evaluated numerically. 
The connected two-point 
correlation function ${\cal C}_r (t)\equiv {\cal G}_r(t)-\rho(t)^2$
follows immediately from the above. 
To establish a connection with the results derived previously \cite{Droz}
we explicitly evaluate ${\cal C}_r (t)$ when $J=0$. Taking due attention to the
apparent singularities occuring in this limit, we find
\begin{eqnarray}
\label{3.30}
\displaystyle
{\cal C}_r(t)&=& \frac{1}{2\pi^2}\int_{0}^{\pi} dq \sin{2\theta_q}\sin{qr}
\int_{0}^{\pi}dq' e^{-2\lambda_{q'}t}\cot{\theta_{q'}}\sin{q'r} \nonumber\\
&-& \Big( \frac{1}{\pi}\int_{0}^{\pi}dq \, e^{-2\lambda_qt}\cos{qr}\Big)^2 
- \Big(\frac{1}{\pi} \int_{0}^{\pi}dq\,e^{-2\lambda_{q}t}\cot{\theta_{q}}\sin{qr}  
\Big) 
\Big(\frac{1}{\pi} \int_{0}^{\pi}dq \, e^{-2\lambda_{q}t}\sin^2{\theta_{q}}\cot{\theta_{q}}\sin{qr}\Big)
\end{eqnarray}
or in terms of elementary functions,
\begin{eqnarray}
\label{3.30bis}
\displaystyle
{\cal C}_r(t)
&=& \sum_{0\leq n<r} e^{-4Dt}\Big(I_{2n-r+1}(4Dt)-I_{2n-r}(4Dt) \Big) \nonumber\\
&+& \Big(\sum_{0\leq n<r}e^{-4Dt} I_{2n-r}\Big)^2 - \Big(\sum_{0\leq n<r}e^{-4Dt} I_{2n-r+1}\Big)^2
- \Big(e^{-4Dt}I_r(4Dt)\Big)^2
\end{eqnarray}
This expression is equivalent to the result derived by a well-known mapping of Lushnikov's model 
to Glauber's $1D$ Ising model \cite{Droz}.\\
Finally, we compute (using formula (\ref{A.15})),  
the asymptotic behaviour of the two-point correlation
function ${\cal C}_r(t)$ when $Dt \gg 1$,\,$Jt\gg 1$, and $r<\infty$
\begin{eqnarray}
\label{3.31}
\displaystyle
{\cal C}_r (t)\sim -\frac{\zeta^{r-1} e^{-4Jt}}{8Jt\sqrt{8\pi Dt}}\Biggr\{ (J+2D)(1-\zeta)^2 +
J(1+\zeta)^2 + \frac{rJ(J+2D)}{D}(1-\zeta^2) 
  \Biggr\} \frac{1}{4\sqrt{J(J+2D)}}
\end{eqnarray}
As expected for a finite source intensity $J>0$, the density and one-time correlation
function decay exponentially with time. Morever, the one-time correlation function also decays
exponentially with distance.\\
A similar calculation can be performed for the case of an initial empty lattice
 ($|\phi_0 \rangle \equiv |0\rangle$). In this case the density is
  $\rho(t)=2J \int_{t}^{\infty} dt'\Big(I_0(4Dt')+I_1(4Dt')\Big)e^{-4(J+D)t'}\sim 
  \rho_{eq}-e^{-4Jt}/\sqrt{8\pi Dt}$ {\cite {Lushnikov}}. It has been shown \cite{Stinch1}, via the mapping to the
  Glauber dynamics, that the nearest neighbor connected function decays ($ Dt,Jt\gg 1$)
  as $C(r=1,t)\sim \rho_{eq}e^{-4Jt}/ \sqrt{2\pi Dt}$.\\
The large-time and large-distance behavior of the one-time correlation
function can also be studied. In fact, in the {\it critical} case it has been shown \cite{Family,Droz} that the 
one-time correlation function obeys a scaling form.\\
Here we investigate the behavior of ${\cal C}_r(t)$ in the limit where $r$ and $Dt \rightarrow \infty$
with $u\equiv \frac{r^2}{8Dt}$  finite. In the {\it massive} case,
we do not expect a scaling form :
\begin{eqnarray}
\label{3.32}
\displaystyle
{\cal C}_r (t)\sim -\sqrt{1+\frac{2D}{J}}\zeta^{r-1}(1-\zeta^2)\frac{g(r^2, u)}{4r^2},
\end{eqnarray}
where
\begin{eqnarray}
\label{3.33}
\displaystyle
g(r^2, u)\equiv \sqrt{\frac{u^3}{\pi}}e^{-\frac{Jr^2}{2Du}}
\end{eqnarray}
The effect of the source  is to disrupt the spatial correlations, i.e. to make
them short-range. In this sense the finite source prohibits
the formation of arbitrarly large vacancy domains.
\\

In the next section we will introduce a field-theoretical approach to deal with 
two-time correlation function. Our approach and the results obtained thereby complement the
exact treatments of the critical case \cite{Stinch3}.
It is worth pointing out, that Glauber calculated in his pioneering work the two-time two-points 
of spin correlations functions\cite{Glauber,Bray}.
\section{Field theoretical Reformulation}
 
The purpose of this section is to reformulate the results of the previous 
section in a field theoretical language, i.e., in terms
of path integrals of fermionic variables (see for example \cite{Negele}). We define the Grassmann numbers $\eta_{q},\eta_{q}^{\ast}$ 
which anticommute with each 
other and with the fermionic operators introduced in the previous section, i.e., 
$\{\eta_q,\eta_{q'}\}=\{\eta_q^{\ast},\eta_{q'}^{\ast}\}
=\{\eta_q^{\ast},\eta_{q'}\}= 0$ and
$\{\eta_q,\xi_{q'}\}= 
\{\eta_q^{\ast},\xi_{q'}^+\}=\{\eta_q^{\ast},
\xi_{q'}\}=\{\eta_q^{\ast},\xi_{q'}^+\}= 0$.
We follow standard practice and
consider the {\it coherent states} associated to the fermionic variables.
We recall that pseudo-fermionic operators required the introduction 
of a {\it left vacuum}  
$\langle \widetilde \chi|$ and a {\it right vacuum} 
$| \chi\rangle$, ergo we define the right  
$|\eta\rangle$ and the left coherent states 
$ \langle \widetilde \eta|$, respectively, as $|\eta\rangle =
e^{-\sum_q \eta_q \xi_q^{+}}|\chi\rangle\;\;\;,\;\;\;\langle \widetilde\eta| =e^{-\sum_q \xi_q\eta_q^{\ast}
}$\\
Despite the fact that $\xi $ and $\xi^{+}$ are not adjoint of each other,
we find the familiar results:\\ $\langle \widetilde \eta|\chi\rangle=\langle \widetilde \chi|\eta\rangle=1\;\;\;,\;\;\;
\langle \widetilde \eta|\eta\rangle=e^{\sum_q \eta_q^{\ast}\eta_q}$ and 
$\xi_q|\eta\rangle=\eta_q|\eta\rangle \;\;,\;\;\; \langle \widetilde 
\eta|\xi_q=\frac{\partial}{\partial \eta_q^{\ast}}\langle\widetilde
\eta| \;\;,\;\;\;
\langle \widetilde \eta| \xi_q^{+}= \langle \widetilde \eta|\eta_q^{\ast}\;\;,\;\;\; 
\xi_q^{+}|\eta\rangle=-\frac{\partial}{\partial \eta_q}
|\eta\rangle $.\\
Most importantly, the closure relation holds true, i.e.,
\begin{eqnarray}
\label{4.7.0}
\displaystyle
\int \prod_{q}d\eta_q^{\ast}d\eta_q e^{-\sum_q \eta_q^{\ast}\eta_q} |\eta\rangle 
\langle \widetilde \eta | =\openone
\end{eqnarray}
At this point, we know from field theory 
how to calculate $\langle \widetilde \chi| n_m e^{{\cal L} 
t}|all\rangle$ and 
$\langle \widetilde \chi| n_m n_{m+r} e^{{\cal L} t}|all\rangle$ using 
the path-integral formalism. 
We discretize time in $M$ infinitesimal intervals
of width $\epsilon \equiv \lim_{M\rightarrow \infty} \frac{t}{M}$. ${\cal L}$ is normal
 ordered, and the closure relation is inserted into the above formulas. We have for the density

\begin{eqnarray}
\label{4.7}
\displaystyle
\rho(t)=\int \prod_{q,\alpha=1\ldots M}d\eta_{q,\alpha}^{\ast}d\eta_{q,\alpha} e^{-\sum_{q,\alpha}
 \eta_{q,\alpha}^{\ast}\eta_{q,\alpha}}   
\langle \widetilde \chi| n_m|\eta_M\rangle
\langle\widetilde\eta_M|e^{{\cal{L}}\epsilon_M}|\eta_{M-1}\rangle \ldots 
\langle\widetilde\eta_2|e^{{\cal{L}}\epsilon_2}|\eta_{1}\rangle
\langle\widetilde\eta_1|e^{{\cal{L}}\epsilon_1}|all\rangle \;\;,
\end{eqnarray}
where
\begin{eqnarray}
\label{4.8}
\displaystyle
\lim_{M\rightarrow \infty} \langle\widetilde\eta_1|e^{{\cal{L}}\epsilon_1}|all\rangle=
e^{-\frac{1}{2}\sum_{q}\cot{\theta_{q}} \eta_{1,q}^{\ast}\eta_{1,-q}^{\ast}  } 
\\
 \langle \widetilde \chi| n_m|\eta_M\rangle=\frac{1}{N}\Big(\sum_{k}\sin^2{\theta_k}-\sum_{k,k'}\sin{\theta_k}
\cos{\theta_{k'}}
 \langle \widetilde \chi| \xi_{-k}\xi_{k'}|\eta_m\rangle\Big)\nonumber\\
=\frac{1}{N}\Big(\sum_{k}\sin^2{\theta_k}-\sum_{k,k'}\sin{\theta_k}\cos{\theta_{
k'}}\,
\, \eta_{M,k} \eta_{M,k'} \Big)
\end{eqnarray}
Taking the continuum limit \cite{Negele}, we arrive at
\begin{eqnarray}
\label{4.9}
\displaystyle
\rho(t)=\frac{1}{N}\int \prod_{q}d\eta_q^{\ast}(t)d\eta_q(t) 
\Big(\sum_{k}\sin^2{\theta_k}-\sum_{k,k'}\sin{\theta_k}\cos{\theta_{k'}}\,\,
\eta_{q}(t) \eta_{q'}(t) 
\Big)e^{-\sum_{q}\frac{\cot{\theta_q}}{2}\eta_q^{\ast}(0)\eta_{-q}^{\ast}(0)} 
e^{-S_0(\eta^{\ast}(t),\eta(t))} \;\;,
\end{eqnarray}
where, the Liouvillian operator 
${\cal L}$ acts as the {\it Hamiltonian} of Field theory. 
By abuse of language we call
$S_0(\eta^{\ast}(t),\eta(t))$ the {\it Euclidian action} of our 
problem, which is defined by
\begin{eqnarray}
\label{4.10}
\displaystyle
S_0(\eta^{\ast}(t),\eta(t))\equiv \int_{0}^{t}dt'\,\, 
\sum_{q}\Big(\eta_q^{\ast}(t')\partial_{t'} \eta_q (t') -
{\cal 
L}(\eta_q^{\ast}(t'),\eta_q(t'))\Big)+\sum_{q'}\eta_{q'}^{\ast}(0)\eta_{q'}(0)
\end{eqnarray}
Taking averages $\langle \ldots \rangle_{S_0}$ on the gaussian distribution, the density
can be rewritten as
\begin{eqnarray}
\label{4.11}
\displaystyle
\rho(t)=\frac{1}{N}\Big \langle 
\Big(\sum_{k}\sin^2{\theta_k}-\sum_{k,k'}\sin{\theta_k}\cos{\theta_{k'}}\,\,
\eta_{q}(t) \eta_{q'}(t) 
\Big)e^{-\sum_{q}\frac{\cot{\theta_q}}{2}\eta_q^{\ast}(0)\eta_{-q}^{\ast}(0)} 
\Big \rangle_{S_0}
\end{eqnarray}
We proceed to discretize 
the free Euclidian action $S_0$ which is bilinear, as
\begin{eqnarray}
\label{4.12}
\displaystyle
S(q)=
\left(
 \begin{array}{c c c c c c}
 1 & 0 & 0 & \ldots & \ldots & 0 \\
 -a & 1 & 0 & \ldots & \ldots & 0 \\
 0 & -a & 1 & 0 & \ldots&0 \\
 \vdots &\ddots &\ddots &\ddots &\vdots &\vdots\\
 0 &\ddots &\ddots &\ddots &\vdots &0 \\
 0 &\ldots &\ldots &0 &-a &1
\end{array}\right)\
\end{eqnarray}
where, $a\equiv 1-\frac{t}{M} \lambda_q$.\\
Following Ref. \cite{Negele}, we find  $\det(S(q))=1$
and 
\begin{eqnarray}
\label{4.11.1}
\displaystyle
\lim_{M\rightarrow \infty} 
S^{-1}_{\alpha,\beta}(q)=e^{-\lambda_q(t_{\alpha}-t_{\beta})},\,\, 
\forall \alpha = 1\ldots M \geq \beta,\,\,\,\, t_{\alpha}=\frac{\alpha t}{M}
\end{eqnarray}
Futhermore we have, in the continuum limit 
$\Bigg(\int\prod_{q} d\eta_{q}^{\ast}(t)d\eta_{q}(t) \,\, e^{-S_0} =1\Bigg)$
\begin{eqnarray}
\label{4.13}
\displaystyle
\langle \eta_p(t_2) \eta_{p'}^{\ast}(t_1)\rangle_{S_0}=\int 
\prod_{q}d\eta_{q}^{\ast}(t)d\eta_{q}(t)\,\, \eta_p(t_2)\eta_{p'}^{\ast}(t_1) e^{-S_0}=
\delta_{p,p'}e^{-\lambda_p(t_2-t_1)}
\end{eqnarray}
Applying Wick's theorem, leads to
\begin{eqnarray}
\label{4.14}
\displaystyle
\langle \xi_{p} \xi_{p'}\rangle (t)= \Big \langle \eta_{p}(t) \eta_{p'}(t) e^{-\frac{1}{2}\sum_{q} \cot{\theta_q}
\eta_{q}^{\ast}(0) \eta_{-q}^{\ast}(0)} \Big \rangle_{S_0} =\delta_{p,-p'}\cot{\theta_p}e^{-2\lambda_p t}
\end{eqnarray}
With this result, it is straightforward to calculate the density using (\ref{4.14}) and (\ref{4.11}).

The computation of correlation function requires evaluation of terms like
 $\langle \xi_{k1} \xi_{k2} \xi_{k3} \xi_{k4} \rangle (t) $ . We  generate in the standard
 fashion the multi-point correlation functions. The generating functional in discretized time reads:
\begin{eqnarray}
\label{4.15}
\displaystyle
Z[g_{q,\alpha}; g^{\ast}_{q',\alpha'}]=\Biggr\langle 
e^{\sum_q\Big\{-\frac{1}{2}\cot{\theta_q}\eta_{q,1}^{\ast} 
\eta_{-q,1}^{\ast} + \sum_{\alpha\geq 2}(g_{q,\alpha}^{\ast}\eta_{q,\alpha} + 
\eta_{q,\alpha}^{\ast} g_{q,\alpha} ) \Big\}}
\Biggr\rangle_{S_0}
\end{eqnarray}
here $g_{q,\alpha},g_{q,\alpha}^{\ast} $ denote (Grassman-numbers) the
coefficient of the source 
terms. Note that this functional differs from the usual field theoretical one
\cite{Negele} by a term that codes the initial condition, i.e.,
$\prod_{q>0}e^{\cot{\theta_q}\eta_{q,1}^{\ast}\eta_{-q,1}^{\ast}}$.
Taking this term into account however is no trouble, i.e.,  
\begin{eqnarray}
\label{4.16}
\displaystyle
Z[g_{q,\alpha}; 
g^{\ast}_{q',\alpha'}]=\Biggr\{\prod_{q>0}\Big(e^{g_{-q,1}^{\ast} 
g_{q,1}^{\ast}\cot{\theta_q}} 
+  g_{-q,1}^{\ast} g_{q,1}^{\ast}g_{-q,1} g_{q,1}\Big)
\Biggr\}\prod_{q,q'}\prod_{i,j=2}^{M}e^{g_{q,i}^{\ast}S_{i,j}(q)^{-1}g_{q',j}\delta_{q,q'}}
\end{eqnarray}
Considering that the source term will be set to zero in the end of the
calculation and noting that 
for $\alpha=1 $, only pseudo-creation operators contribute, 
we find it convenient to work with
\begin{eqnarray}
\label{4.17}
\displaystyle
\widetilde Z[g_{q,\alpha}; g^{\ast}_{q',\alpha'}]=\prod_{q}e^{\Big(\frac{1}{2} 
g_{-q,1}^{\ast}
g_{q,1}^{\ast}\cot{\theta_q}+\sum_{q'}\sum_{i,j>1}g_{q,i}^{\ast}e^{\lambda_q 
(t_j-t_i)}g_{q',j} \delta_{q,q'}   \Big)}
\end{eqnarray}
From the definition, it follows then that
\begin{eqnarray}
\label{4.18.0}
\displaystyle
\langle\xi_{k_1}\xi_{k_2}\rangle (t)=\frac{ \delta^6 \widetilde 
Z[g_{q,\alpha};g_{q',\alpha'}^{\ast}]}{\delta g_{-q,1}
\delta g_{q,1}^{\ast}  \delta g_{k_1,M}^{\ast} \delta g_{k_2,M}^{\ast} \delta g_{q,2} 
\delta g_{-q,2}}\Biggr|_{g=g^{\ast}=0}
\end{eqnarray}
or more generally,
\begin{eqnarray}
\label{4.18.1}
\displaystyle
\langle\xi_{k_1}\xi_{k_1'}\ldots \xi_{k_n}\xi_{k_n'} \rangle (t)\equiv
\langle \widetilde \chi| \xi_{k_1}\xi_{k_1'}\ldots \xi_{k_n}\xi_{k_n'} e^{{\cal L}t}|all \rangle =
 \langle \widetilde \chi| \xi_{k_1}\xi_{k_1'}\ldots \xi_{k_n}\xi_{k_n'} e^{{\cal L}t}e^{-\frac{1}{2}\sum_q{\cot{\theta_q}
 \xi_q^{+}\xi_{-q}^{+}}}|\chi \rangle 
\nonumber\\=
\frac{ \delta^{6n} \widetilde Z[g_{q,\alpha};g_{q',\alpha'}^{\ast}]}{\delta 
g_{-q_n,1}^{\ast} 
\delta g_{q_n,1}^{\ast} \ldots  \delta g_{-q_1,1}^{\ast} \delta g_{q_1,1}^{\ast} 
\delta g_{k_1,M}^{\ast} \delta g_{k_1',M}^{\ast}  \ldots \delta g_{k_n,M}^{\ast} \delta g_{k_n',M}^{\ast} 
\delta g_{q_1,2} \delta g_{-q_1,2} \ldots  \delta g_{q_n,2} 
\delta g_{-q_n,2}^{\ast}} \Biggr|_{g=g^{\ast}=0}
\end{eqnarray}
As an application of this field-theoretical formulation we will now extend  our approach
to the computation of the two-time correlation function, for an initial state $|\phi_0\rangle$,

\begin{eqnarray}
\label{4.19}
\displaystyle
{\cal G}_r(t,t')\equiv \langle n_{l+r}(t+t') n_l (t) \rangle =\langle \widetilde \chi| n_l e^{{\cal L}t'} n_{l+r} e^{{\cal
L}t}|\phi_0 \rangle  =
\sum_{n,n'}\langle n'|n_{l+r} e^{{\cal L}t'} n_l e^{{\cal L}t} |n \rangle  P(n,0) \nonumber \\
= \sum_{n,n',n''}\langle n'|n_{l+r} |n'\rangle  \langle n''|n_{l} |n''\rangle W_{n'n''}(t') W_{n''n}(t) P(n,0)
\end{eqnarray}
As above, we consider an initially filled lattice ($|\phi_0\rangle = |all \rangle$).\\
To perform the computation of ${\cal G}(r,t,t')$ we need to evaluate terms like 
\begin{eqnarray}
\label{4.21}
\displaystyle
\langle \widetilde \chi | \xi_p \xi_{p'} e^{{\cal L}t'}\xi_q \xi_{q'} e^{{\cal L}t}|all \rangle =
\Big \langle \eta_p (t+t') \eta_{p'} (t+t') \eta_q (t) \eta_{q'} (t) e^{-\frac{1}{2}\sum_k \cot{\theta_k}
\eta_k^{\ast}(0)\eta_{-k}^{\ast}(0) }  \Big \rangle_{S_0}
\end{eqnarray}
This expression can be calculated by using the generating functional, where as before $t$
 ($t'$) is discretized into $M$ ($N$) infinitesimal
time steps (respectively), with $M,N \rightarrow \infty$:
\begin{eqnarray}
\label{4.22}
\displaystyle
\langle \widetilde \chi | \xi_p \xi_{p'} e^{{\cal L}t'}\xi_q \xi_{q'} e^{{\cal L}t}|all \rangle &=& 
\frac{ \delta^{12} \widetilde Z[g_{q,\alpha};g_{q',\alpha'}^{\ast}]}{\delta g_{-q_2,1}^{\ast} 
\delta g_{q_2,1}^{\ast} \delta g_{-q_1,1}^{\ast} \delta g_{q_1,1}^{\ast} 
\delta g_{q,M}^{\ast} \delta g_{q',M}^{\ast} \delta g_{p,M+N}^{\ast} \delta g_{p',M+N}^{\ast}
\delta g_{q_1,2} \delta g_{-q_1,2} \delta g_{q_2,2} 
\delta g_{-q_2,2}  } \Big|_{g=g^{\ast}=0}\nonumber\\
&=& e^{-(\lambda_p + \lambda_{p'})(t+t')-(\lambda_q + \lambda_{q'})t}
\Biggr (\cot{\theta_p}\cot{\theta_q}\delta_{p,-p'}\delta_{q,-q'} \nonumber\\ &+& \cot{\theta_p}\cot{\theta_{p'}}
\Big(\delta_{-p,q'}\delta_{-q,p'} - \delta_{-q,p}\delta_{-p',q'} \Big)\Biggr)
\end{eqnarray}
where  the continuum limit for the time has been taken.\\
Applying the same technique to each term of ${\cal G}(r,t,t')$, yields:
\begin{eqnarray}
\label{4.23}
\displaystyle
{\cal G}_r(t,t')&=&\rho_{eq}^2 +2(J+2D)\rho_{eq}\int_{t}^{\infty}dT e^{-4(J+D)T}\Big(I_0(4DT)-
I_1(4DT)\Big)\nonumber \\
&+& J(J+2D)\Biggr(\int_{t'}^{\infty} dT e^{-2(J+D)T}\Big\{I_r(2DT)+I'_r(2DT)\Big\}\Biggr) 
 \nonumber \\ &\times &
\Biggr(\int_{t'}^{\infty} dT e^{-2(J+D)T}\Big\{I_r(2DT)- I'_r(2DT)\Big\}\Biggr)\nonumber \\
&+& \frac{J(J+2D)}{4}\Biggr(\int_{t'}^{\infty}dT e^{-2(J+D)T} \frac{r}{DT}I_r (2DT)\Biggr)^2  \nonumber\\
&-& 2(J+2D)^2 \Biggr( \int_{t'}^{\infty}dT e^{-2(J+D)T}
\Big\{I_r(2DT)-I'_r(2DT)\Big\}\Biggr)
 \nonumber \\ &\times &
\Biggr(\int_{t}^{\infty} dT e^{-2(J+D)(2T+t')}\Big\{I_{r}(2D(2T+t'))-I'_{r}(2D(2T+t'))\Big\}
 \Biggr)\nonumber\\
&+& (J+2D)^2\Biggr( \int_{t'}^{\infty} dT 
e^{-2(J+D)T}\frac{r}{2D(2T+t')}I_r(2DT) \Biggr)
 \nonumber \\ &\times &
\Biggr(\sum_{0\leq n<r}\int_{t}^{\infty} dT e^{-2(J+D)(2T+t')}\Big\{2I_{2n-r+1}(2D(2T+t'))-2I_{2n-r}(2D(2T+t'))
+ I'_{2n-r}(2D(2T+t'))\Big\}\Biggr) \nonumber \\
&+& J(J+2D) \Biggr(\int_{t}^{\infty} dT e^{-2(J+D)(2T+t')} \Big(I_0(2D(2T+t'))- I_1(2D(2T+t'))\Big)\Biggr) \nonumber \\
& \times & \Biggr(\int_{t'}^{\infty}dT e^{-2(J+D)T} \Big(I_0(2DT)+ I_1(2DT)\Big)\Biggr)
\nonumber \\ &-& 
\frac{J(J+2D)}{2}\Biggr(\int_{t}^{\infty}dT e^{-2(J+D)(2T+t')} 
\frac{r}{D(2T+t')}I_{r}(2D(2T+t'))\Biggr) \Biggr(\int_{t'}^{\infty}dT e^{-2(J+D)T} \frac{r}{DT}I_r(2DT)\Biggr)\nonumber\\
&-& 2J(J+2D) \Biggr(\int_{t'}^{\infty}dT e^{-2(J+D)T}
\Big\{I_r (2DT)+I'_r (2DT)\Big\}\Biggr) \nonumber \\ &\times&
\Biggr(\int_{t}^{\infty}dT e^{-2(J+D)(2T+t')}\Big\{I_{r}(2D(2T+t'))- I'_{r}(2D(2T+t'))\Big\}\Biggr) 
\nonumber \\ 
&+& 4(J+2D)^2 \Biggr(\int_{t}^{\infty}dT e^{-2(J+D)(2T+t')} \Big(I_0(2D(2T+t'))- I_1(2D(2T+t'))\Big) \Biggr)^2 
\nonumber \\
&-& 2(J+2D)^2 \Biggr(\int_{t}^{\infty}dT e^{-2(J+D)(2T+t')} \Big\{ I_{r}(2D(2T+t'))-I'_{r}(2D(2T+t'))
\Big\}\Biggr)^2 \nonumber\\
&+&2(J+2D)^2 \Biggr(\sum_{0\leq n<r}(2n-r)\int_{t}^{\infty} dT\, e^{-2(J+D)(2T+t')}
\frac{I_{2n-r}(2D(2T+t')) }{D(2T+t')}\Biggr)  \nonumber \\
&\times & \Big(\sum_{0\leq n<r}\int_{t}^{\infty} dT\, e^{-2(J+D)(2T+t')}
\{2I_{2n-r+1}(2D(2T+t')- 2I_{2n-r}(2D(2T+t'))
+I'_{2n-r}(2D(2T+t'))\}\Big)
\end{eqnarray}
In the critical case some care is required when taking $J \rightarrow 0$,
\begin{eqnarray}
\label{4.24}
\displaystyle
{\cal G}_r(t,t')&=& e^{-4D(2t+t')}\Biggr\{\Big(I_0(2D(2t+t'))\Big)^2 - \Big(I_r(2D(2t+t'))\Big)^2\Biggr\} \nonumber\\ &+&
De^{-2D(2t+t')}\Biggr\{ \int_{t'}^{\infty} dT e^{-2DT}\frac{r}{DT} I_{r}(2DT) \Biggr\} \nonumber\\
&\times & \Biggr\{\sum_{0\leq n <r} \Big(I_{2n-r+1}(2D(2t+t')) - I_{2n-r}(2D(2t+t'))\Big)   \Biggr\} \nonumber \\ &+&
e^{-4D(2t+t')}\Biggr\{ \Big(\sum_{0\leq n<r} I_{2n-r}(2D(2t+t'))\Big)^2 -  \Big(\sum_{0\leq n<r} I_{2n-r+1}(2D(2t+t'))\Big)^2 \Biggr\}
\end{eqnarray}
To study the behavior at finite distance of the two-time correlation function ${\cal C}_r(t,t')\equiv{\cal G}_r(t,t') -\rho(t)\rho(t')$ 
 , we distinguish  the massive and the critical regime.
We begin with the massive case and  consider $Dt,Dt' \gg 1$, $Jt, Jt'\gg 1$  and $r< \infty$. 
 With help of (\ref{A.12}-\ref{A.15}), we have:
\begin{eqnarray}
\label{4.25.0}
\displaystyle
{\cal G}_r(t,t') &\sim & \rho_{eq}^2 + \Big(1+\frac{J}{2D}\Big)\rho_{eq}\frac{e^{-4Jt}}{8Jt \sqrt{8\pi Dt}}\nonumber \\
{\cal C}_r(t,t') &\sim & - \Big(1+\frac{J}{2D}\Big)\rho_{eq} \frac{e^{-4Jt'}}{8Jt'\sqrt{8\pi Dt'}}
\end{eqnarray}
When $r < \infty,\,Dt$ and $Jt$ are finite and $Dt', Jt' \gg 1$ we obtain: 
\begin{eqnarray}
\label{4.25.1}
\displaystyle
{\cal G}_r(t,t') &\sim & \rho_{eq}^2 + \frac{J+2D}{128J}\frac{e^{-4Jt'}}{(Dt')^2}+ \frac{J+2D}{4JD}
\frac{e^{-4J(t+t')}}{16 \pi D \sqrt{t'}(2t+t')^{\frac{3}{2}}}\nonumber \\
{\cal C}_r(t,t') &\sim & -\Big(1+\frac{J}{2D}\Big)\rho_{eq} \frac{e^{-4Jt'}}{8Jt' \sqrt{8\pi Dt'}}
\end{eqnarray}
In the critical case ($r<\infty$), both asymptotic  behavior $Dt, Dt'\gg 1$ and 
 $Dt$ finite with $Dt'\gg 1$ of the disconnected and connected 
correlation function are given by:
\begin{eqnarray}
\label{4.26}
\displaystyle
{\cal G}_r(t,t') &\sim & \frac{r^2 }{4\pi D^2
(2t+t')^2}\Big(1+\frac{1}{8}\sqrt{1+\frac{2t}{t'}}\Big)  \nonumber \\
{\cal C}_r(t,t') &\sim & -\frac{1}{8\pi D \sqrt{t t'}}
\end{eqnarray}
\\
Near the initial state (the density of particles is high), when $Dt, Dt' \ll 1$  the decay is linear 
 and independ of $r$, i.e., 
\begin{eqnarray}
\label{4.28}
\displaystyle
{\cal G}_r(t,t') &\sim & 1-4D(2t+t')  \nonumber \\
{\cal C}_r(t,t') &\sim & -4Dt
\end{eqnarray}
We now provide a scaling form for the two-time correlation function ${\cal C}_r (t,t')$.
It is known, from the duality with Glauber's model, that the single-time correlation function obeys a 
scaling form for large-time and long-distances ($Dt$ and 
$ r \rightarrow \infty$ with the  ratio $r^2/Dt$ held finite \cite{Family,Droz}).
The scaling form is found as ${\cal C}_r(t)\sim r^{-2} f(r^{2}/4Dt)$, where the exponent $-2$ 
is believed to be universal \cite{Droz}.
We further assume the long-time and large-distance scaling form  
${\cal C}_r(t,t')\sim r^{-y} h(u,v)$, where $r$, $Dt$, $Dt' \rightarrow \infty $ with
$u\equiv r^2/4Dt$ and $v\equiv r^2/4Dt'$ held finite and arrive at
\begin{eqnarray}
\label{4.29}
\displaystyle
{\cal C}_r(t,t') &\sim & \frac{1}{\pi r^2}\Biggr\{K^{\frac{3}{2}}e^{-K}
\Big( \sqrt{\pi}\, {\rm erf}{\sqrt{v}} - 2\sqrt{K}\Big) +K\Big(1-e^{-2K}\Big)
-\frac{\sqrt{uv}}{2}\Biggr\},
\end{eqnarray}
where  $0< K\equiv \frac{uv}{u+2v} < \infty$.
\\
At this point, it is appropriate to review  what we have achieved so far: 
we have been able to reformulate the problem of the evaluation of the 
multi-point
correlation functions in a language that parallels the field theoretical one.
This allows us to compute in an efficient and systematic way physical quantities
of interest despite some technical differences to the standard approach. 
While 
this paper deals with a free ``field theory'' of pseudo-fermions, it is tempting
 to apply the same formalism to the multi-species case
where two- or multi-body interactions arise. The latter however will be
investigated in a future work 
by pertubative renormalization group techniques, as no exact
solutions are available. 
\section{Concluding remarks}
We have studied three different approaches to the
problem of  diffusion-annihilation of
classical hard-core particles moving on
a one-dimensional ring. 
Though Lushnikov's contributions to the problem are genuine and undisputable, 
we have
shown how an extension of his generating function method to evaluate the
two-point correlation function can be cumbersome in
practice, even in the simplest case available of a single-species. 
We have seen that it is advantageous to apply a
 generalized Bogoliubov transformation first used by Grynberg et {\it al.}
\cite{Stinch1,Stinch2} in a different context. The evolution operator 
can be diagonalized, i.e., expressed as a quadratic form of two operators 
that are not adjoint of each other. Despite this fact
the formalism resembles the standard one and
appears as a powerful tool. Indeed, we were able
to compute for the first time the full one-time and  two-time correlation functions 
 for an initially fully occupied lattice
(other initial conditions can also be studied) in the presence of a finite
source. We derived a scaling form for the two-time correlation function.
We used the results of the first section (algebraic decay in the 
critical case and exponential in the presence of source) to check the
asymptotic behavior of the density and two-point correlation function.
We discovered that while in the absence of source, the modes at long wave 
lengths
fully control the long-time asymptotics of the density and correlation
functions, in the presence of a finite source all modes contribute. This means
that in the general case, the long wave-length approximations that were so
successfull in strongly
correlated systems (such as bosonization or conformal field theory techniques) 
do not work for the problem at hand.
Moreover, the idea of exploiting the integrability of some spin Hamiltonians 
on which the
multi-species Liouvillian maps, might turn out more elusive than expected.
In view of the above remark, it would seem extremely difficult if not
impossible to
extract from the exact Bethe-ansatz solution of the non-hermitian
spin Hamiltonian the relevant matrix
elements that in turn allow the evaluation of correlation functions \cite{Droz}.   
We propose to tackle the multi-species problem in terms of fermion
functionals, the main difficulties arising from the two- and/or multi-body
interactions occuring in the process of mapping classical particles to fermions. 
We will illustrate the power of the formalism in a
forthcoming paper, where we will apply the renormalization group scheme.
Besides the obvious advantage of formulating 
the problem in a
field theoretical language (perturbation theory,...), the method is 
applicable to arbitrary densities of particles, and thus
complements the approach developed by Cardy and collaborators \cite{Cardy1,Cardy2,Cardy3,Lee}.
In higher dimensions, however, Fermi statistics requires the introduction of a
gauge field that is strongly coupled to the fermions. We also intend to explore
this line of research in the future.  
%
%
\section*{ACKNOWLEDGMENTS}
We thank M. Droz, L. Frachebourg, Ch. Gruber, S. Gyger, Ph. Nozi\`{e}res and T.M. Rice for stimulating
discussions.
We are grateful to the referee for drawing our attention to references
\cite{Stinch3,Bray}.
The support of the Swiss National Fonds is gratefully acknowledged.
\appendix
\section{Useful results}
In this appendix we provide some useful properties of 
the Bessel functions of imaginary argument. 
We recall the definition of the modified Bessel $I_n(z)$ function
($ n $ integer) \cite{Grad}
\begin{eqnarray}
\label{A.1}
\displaystyle
\ I_n(z)=\frac{(-i)^n}{2\pi}\int_{-\pi}^{\pi} e^{z \sin{(p)}+i \,np} dp 
=\frac{1}{\pi} \int_{0}^{\pi} e^{z \cos{\theta}} \cos{(n\theta)} d\theta
\end{eqnarray}
with $I_n(z)= I_{-n}(z)$. We also use in (\ref{3.27.1},\ref{4.23},\ref{4.24}) the well-known properties
$I_{n-1}(z)-I_{n+1}(z)= \frac{2n}{z} I_{n}(z)$ and 
$I_{n-1}(z)+I_{n+1}(z)= 2\frac{d}{dz} I_{n}(z)$.
The following integrals occur in the evaluation of the two-point 
correlation function.
\begin{eqnarray}
\label{A.4}
\displaystyle
\ \widetilde I_1 (r,t)\equiv \int_{0}^{\pi}\frac{\sin {qr}}{\sin {q}} e^{4Dt 
\cos{q}} dq\;\;,\;\; \widetilde I_2 (r,t) \equiv \int_{0}^{\pi}\frac{\sin {qr}}{\sin {q}} \cos{q}\; 
e^{4Dt \cos{q}} dq\;\;,\;\; \widetilde I_3 (r,t) \equiv \int_{0}^{\pi}\frac{\sin {qr}}{\sin {q}} \cos{2 q}\; 
e^{4Dt \cos{q}} dq
\end{eqnarray}
Setting $\widetilde q = q-i\epsilon, \,
 \epsilon $ real and  $\epsilon>0$, we have
\begin{eqnarray}
\label{A.7}
\displaystyle
\ \frac{\sin {qr}}{\sin {q}}= \lim_{\epsilon \downarrow 0}\frac{\sin {\widetilde q 
r}}{\sin {\widetilde q}}
\end{eqnarray}
\begin{eqnarray}
\label{A.8}
\displaystyle
\ \frac{\sin {qr}}{\sin {q}}= \lim_{\epsilon \downarrow , \widetilde q \rightarrow 
q}\frac{\sin {\widetilde q r}}{\sin {\widetilde q}}= 
\lim_{\epsilon \downarrow
0, \widetilde q \rightarrow q}\sum_{n\geq 0}\Bigr(e^{-2 i \widetilde q 
(n-(\frac{r-1}{2}))}- e^{-2 i \widetilde q (n+(\frac{r+1}{2}))} \Bigr)
\end{eqnarray}
Therefore,
\begin{eqnarray}
\label{A.9}
\displaystyle
\ \widetilde I_1(r,t)=\int_{0}^{\pi} \Biggr(\lim_{\epsilon \downarrow 0 ,
 \widetilde q \rightarrow q}\sum_{n\geq 0}\Bigr(e^{-2 i \widetilde q 
(n-(\frac{r-1}{2}))}- e^{-2 i \widetilde q (n+(\frac{r+1}{2}))} \Bigr)\Biggr) e^{4Dt 
\cos{q}} dq =
 \nonumber\\
= \pi \sum_{0\leq n < r} I_{2n-r+1} (4Dt)
\end{eqnarray}
Similarly, we find
\begin{eqnarray}
\label{A.10}
\displaystyle
\widetilde I_2 (r,t)= \frac{\pi}{2} \sum_{0\leq n < r} \Bigr\{ I_{2n-r} (4Dt) +  
I_{2n-r+2} (4Dt)\Bigr\}=
         \pi \sum_{0\leq n < r} I_{2n-r} (4Dt)
\end{eqnarray}
\begin{eqnarray}
\label{A.11}
\displaystyle
 \widetilde I_3(r,t)= \frac{\pi}{2} \sum_{0\leq n < r} \Bigr\{ I_{2n-r-1} (4Dt) +  
I_{2n+r-1} (4Dt) +
 I_{2n-r+3} (4Dt) +  I_{2n+r+3} (4Dt)   \Bigr\}= \nonumber\\ = \pi \sum_{0\leq n 
< r} I_{2n-r-1} (4Dt)
\end{eqnarray}
\\
With help of asymptotic behaviour of Bessel functions and using the properties of the Incomplete
Gamma functions \cite{Grad}, the asymptotic behaviour ($Dt\gg 1$,\,$Jt\gg 1$) of the following
integrals is readily found 
\begin{eqnarray}
\label{A.12}
\displaystyle
\int_{t}^{\infty} dt' \;\frac{e^{-4Jt'}}{\sqrt{8\pi Dt'}}\sim 
\frac{e^{-4Jt}}{4J\sqrt{8\pi
Dt}}\Bigr(1-\frac{1}{8Jt}+{\mathcal O}((Jt)^{-2})\Bigr)
\end{eqnarray}
\begin{eqnarray}
\label{A.13}
\displaystyle
\int_{t}^{\infty} dt' \;\frac{e^{-4Jt'}}{\sqrt{8\pi Dt'}8Dt'}\sim 
\frac{e^{-4Jt}}{4J\sqrt{8\pi Dt}}\Bigr(\frac{1}{8Dt}-\frac{3}{64Jt}+{\mathcal O}((Jt)^{-3})\Bigr)
\end{eqnarray}
\begin{eqnarray}
\label{A.14}
\displaystyle
\int_{t}^{\infty} dt' \;\frac{e^{-4Jt'}}{\sqrt{8\pi Dt'}(Dt')^2}\sim 
\frac{e^{-4Jt}}{4J\sqrt{8\pi Dt}}
\Bigr(\frac{1}{(Dt)^2}+{\mathcal O}((Jt)^{-3})\Bigr)
\end{eqnarray}
A futher result ($Dt\gg 1$,\,$Jt\gg 1$) used in the evaluation of the asymptotic 
behaviour of the density and 
the two-point correlation function (\ref{2.22},\,\ref{3.31},\,\ref{4.22},\,\ref{4.23},\,\ref{4.24},\,\ref{4.25.0},\,\ref{4.25.1},\,\ref{4.26})
\begin{eqnarray}
\label{A.15}
\displaystyle
\int_{t}^{\infty}e^{-4(J+D)t'}I_n (4Dt')\sim \frac{e^{-4Jt}}{4J\sqrt{8\pi Dt}}\Biggr\{\Bigr(1-\frac{1}{8Jt}+{\mathcal O}((Jt)^{-2})\Bigr)- \nonumber\\  
-(n^2 - \frac{1}{4})\Bigr(\frac{1}{8Dt}-\frac{3}{64JDt^2}+{\mathcal O}((Jt)^{-3})\Bigr)+ 
\nonumber\\ +(n^2-\frac{1}{4})(n^2-\frac{9}{4})
\Bigr(\frac{1}{128D^2t^2}+ {\mathcal O}((Jt)^{-3}) \Bigr)+\ldots\Biggr\}
\end{eqnarray}
Finally, the calculation of the constants $A_0,B_0, C_0$, when $J>0$ is
performed 
with help of the formula \cite{Grad}
\begin{eqnarray}
\label{A.16}
\displaystyle
\int_{0}^{\infty}dx \; e^{-\alpha x}I_{\nu}(\beta x)= \frac{\beta 
^{\nu}}{\sqrt{\alpha^2 - \beta^2}\Bigr(\alpha+\sqrt{\alpha^2 - \beta^2
}\Bigr)^{\nu}},\,\,\forall \,Re(\nu) >-1,\;\;Re (\alpha)\,> Re|\beta|
\end{eqnarray}
\end{document}